# Framework Architecture for WLAN Testbed

Aphirak Jansang and Anan Phonphoem
Department of Computer Engineering, Faculty of Engineering,
Kasetsart University, Bangkok, 10900, Thailand
aphirak.j@ku.ac.th; anan.p@ku.ac.th

**ABSTRACT**
There has been a tremendous effort in improving wireless LAN for supporting the demanding multimedia application. Many new protocols or ideas have been proposed and proved by using a mathematical model or running a simulation program. That is satisfactory but these proposed designs might not work in the real world situation. Testbed is an option to alleviate this gap and present the opportunity to see the real problem and ensure that the design works. A framework architecture for building a testbed to test a new concept or design is presented in this paper. The framework is designed in the modularity style in such a way that can be easily exchanged or modified. A testbed based on the framework that implements the polling based mechanism has been created and the results have shown that the QoS of the real time traffic can be maintained in the presence of the high non-real time traffic.

## 1. INTRODUCTION

WLAN is now more popular than ever due to its convenience to purchase, install, configure, and even maintain. However, many problems in WLAN required solution for example the limited bandwidth, power consumption and quality of service (QoS) support. Some researchers have proved their concepts and designs by using mathematical models such as numerical analysis and probability or queueing theory models. Others have compared their designs with those by implemented them on proprietary or well-know simulation programs. Nevertheless both methods may be unable to capture the real working situation. By building a real system with a controllable environment called 'Testbed' a trusty method to prove a new protocol, test an algorithm, or observe causes of infant mortality when implement it in the real situation is introduced. To produce a testbed is time and effort consuming but it pays off.

The WLAN testbed that researches normally built is divided into three groups: Firstly by implementing all components (both hardware and software) from ground up, Secondly by implementing only hardware or software modifications on the existing system, and finally by implementing simulation and testbed to work together.

The example of implementing all components from the ground up is entitled TUTWLAN [10-13]. New hardware including drivers for the testbed employing the platform is used Windows NT. The design is in modular fashion which is sufficiently flexible to add or remove each module from the testbed as necessary.

Some researchers [8,16] make their testbed and framework by modifying the protocol stack so that the packet flow can be controlled, by doing this the module becomes independent of both existing hardware and application.

The WHYNET [14-15] is an example of implementing simulation and testbed together. The algorithm or design proved in the simulation can be easily tested by the testbed. The designed framework contains both virtual and real components. The system appears complicated but it is very convenient and effective method for researchers to prove their concepts.

In this paper, a framework for building a WLAN protocol testbed is proposed concentrating on creating a testbed for testing new protocols that supports QoS on the centralized WLAN. The designed framework is independent of both existing hardware and software enabling it to be easily integrated in the real system and also to be modular for easy plugging and unplugging from the existing system. The testbed is based on the Linux platform.

The paper is organized as follows. In the next section, the system framework is proposed. Section 3 shows how to set up the testbed based on the designed framework. Testing method and test results are described in section 4. Finally, the last section summarizes the paper.

## 2. Framework

The basic concept of polling system is now reviewed and then the designed framework which follows the polling based mechanism will be presented.

### 2.1 Polling based media access control

Media access control (MAC) in WLAN are mostly based on CSMA/CA which is the collision based. Both real time and non-real time traffic in the same WLAN segment are treated in the same way. As both real time and non-real time traffic are totally different in their requirements, real time traffic being more sensitive to the delay and delay variation, whereas the non-real time traffic may be more aware of the correctness of the data but not the delay. Voice and video are examples of the real time traffic and for the non-real time traffic the regular Web or Ftp serves.

Many researchers [2-5] have proposed methods to support QoS by implementing some policies based on the centralized polling system [7,9]. This system consists of a central controller, called master station, and other regular stations. The master station controls the media usage by granting permission, called polling, to a regular station (or a connecting session) in the system. Each regular station (session) can transmit data only when it is explicitly permitted by the master station. The bandwidth for each station (session) is allocated based on its QoS requirements. The performance of the system will drop due to the polling overhead especially in the light load situation but the QoS support is worth it to scarify with some overheads.

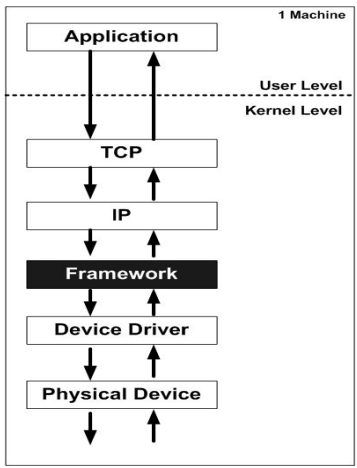

**Figure 1. Framework architecture.**

## 2.2 Packet types
Two principal packet types are used in the system.

- Data packet – a real data packet generated by a source to a destination.
- Control packet – a system generated packet for management purposes consisting of
  - NTS packet – a nothing to send packet. It will be generated when the polled station has no data to transmit after polling.
  - Poll packet – a polling packet generated by the master station to grant permission to any station in the network.

## 2.3 Proposed Framework Architecture
According to design goals, the framework must be independent of both existing hardware and software so that it can be easily integrated into the real system. It is desired to insert the proposed framework module in between the IP and the data link layer [1] (in this case, the device driver layer) as shown in Figure 1. Therefore, the modified system can still work on any existing application or any wireless network interface card. We select the Linux platform due to the open source characteristics. The framework has been designed in a modular style such that the inserted framework part can be easily swapped in and out. The users can elect to invoke or by pass the framework accordingly.

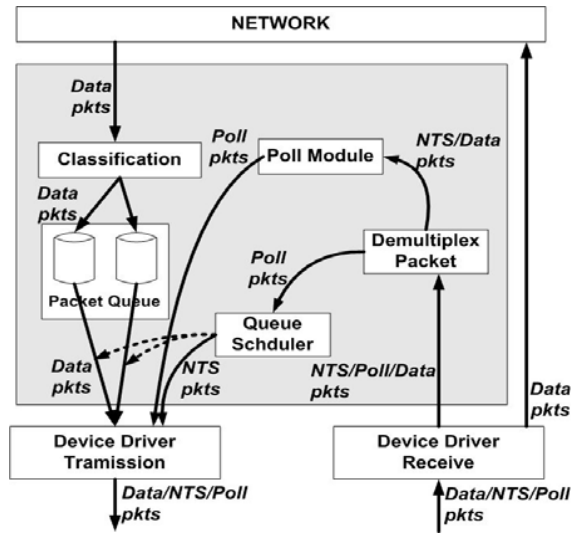

**Figure 2. Framework module in details.**

There are five major components in the framework shown in Figure 2.

- Classification Module – By investigating the header of a sending packet, the module categorizes each packet based of its class, then forwards it to the proper packet queue.
- Packet Queue Module – Each queue represents a traffic class which is independent from the others. The FIFO is implemented as the mechanism for each queue.
- Queue Scheduler Module – Based on the poll packet received from the network, the queue scheduler module will dequeue the packet from the proper queue and transfer it to the device driver transmission module to await transmission.
- Demultiplex Packet Module – By investigating the incoming packet from the network, the demulitplex packet module will send the poll packet to the queue scheduler module for dequeueing certain packets while sending the Nothing To Send (NTS) and Data packet to the poll manager module for monitoring the current situation of the medium.
- Poll Manager Module – This is the principal intelligent part of the framework. Once any new mechanism for controlling the media has been designed, it will be inserted in this module for proving the concept. The poll manager module responses for creating a polling list based on the QoS requirements for each station and the status of the medium. The module also creates polling packets for the system.

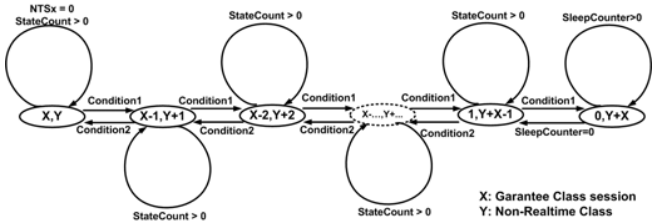

**Figure 3. State transition diagram of the polling adjustment mechanism.**

## 3. Testbed implementation

A testbed having been established based on the proposed framework as shown in the previous section. It must now be orientated to verify any new protocols for QoS support. This is implemented by a simple polling mechanism that can adjust the state of each station according to its load. The state diagram of the polling adjustment is shown in Figure 3 and further details can be found at [17].

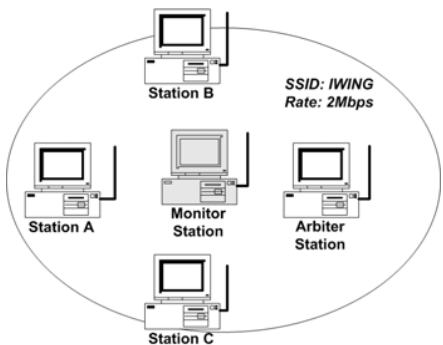

**Figure 4. Testbed setup.**

Figure 4 shows how the testbed has been setup in the Adhoc manner. It is composed of 5 stations: one master station, three regular stations, and one monitoring station. The framework to each station in the network is deployed except for the monitoring station that runs a general sniffer program. Each station is equipped with the IEEE 802.11b wireless network card which is preconfigured to operate in channel 3 at 2 Mbps. All stations are Celeron 400 MHz running the Linux operating system with the kernel version 2.4.18.

## 4. Testing method and test results

In the experiment, all running traffic is classified into two classes: real time and non-real time. Video streams of 250 and 400 Kbps are used for representing real time traffic and ftps represent the non-real time traffic.

The testing method is as followings. Firstly, the system is tested by running a video stream without running any other application. Then increase the number of the ftp sessions one at time and observing quality of the video picture. Meanwhile transferred data among stations has been captured by the monitoring station for performance analysis.

Now run the same test twice for each 250 and 400 Kbps video streams. In the first round, the test is operated without invoking the framework. Later we rerun the system in the same situation and environment with the active framework. The system performance based on the bandwidth, delay and delay variation metrics can now be compared.

Figure 5 shows the results for the experiment of 250 Kbps video stream. In Figures 5a, the jitter can be observed once four ftp sessions are running in the situation of non-active framework. The video session will receive less bandwidth due to the fact that many sessions join the network and compete for the limited bandwidth with equal priority. In the opposite situation, the video quality is maintained when the framework with the polling mechanism is active and bandwidth of the poll-video is also maintained. All ftp sessions are grouped together in the same non-real time queue having lower priority compared to real time traffic, the video session, so each ftp session will experience more delay. However, it does not affect the QoS requirements of the Ftp. In the design, certain bandwidths can be allocated to protect the starvation problem for each traffic class. Figure 5b and 5c also shows the delay and delay variation that each type of traffic can experience. It can be seen that with the active framework the delay and delay variation for the video session is quite stable enough to claim that the QoS for the real time traffic can be guaranteed.

Similar results for the 400 Kbps video can be seen in Figure 6. The only difference is number of ftp sessions that causes the jittering to occur in the non-active framework is quite low due to the limited bandwidth.

## 5. Conclusion

We have proposed the framework architecture that can be deployed to create a testbed. The testbed has been outlined and is useable for testing new MAC protocols or QoS support mechanisms. In this research the framework has been implemented for the centralized polling based media access control system. This has been designed to be independent of the wireless network interface cards and existing applications. The test results have showed that the QoS of the real time traffic has been maintained even in the presence of the high non-real time traffic. Although only two packet queues, or classes, have been implemented, the system has the capability to support multiple classes.

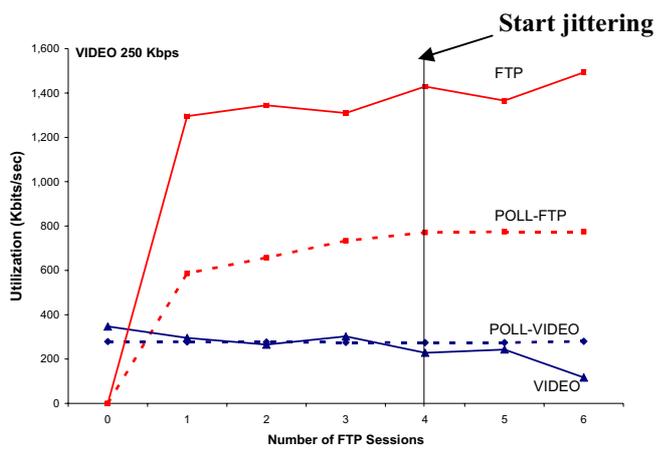

a) Bandwidth utilization

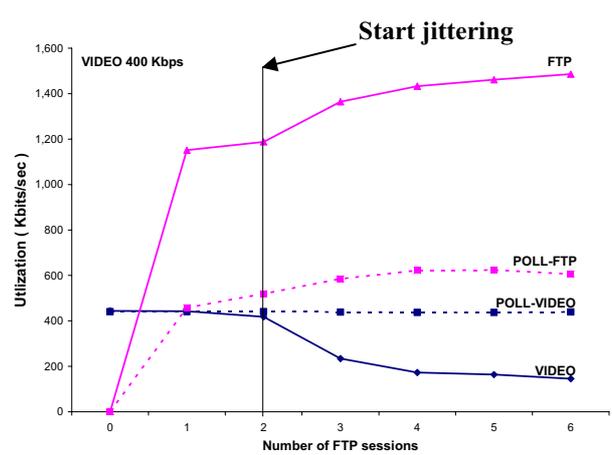

a) Bandwidth utilization

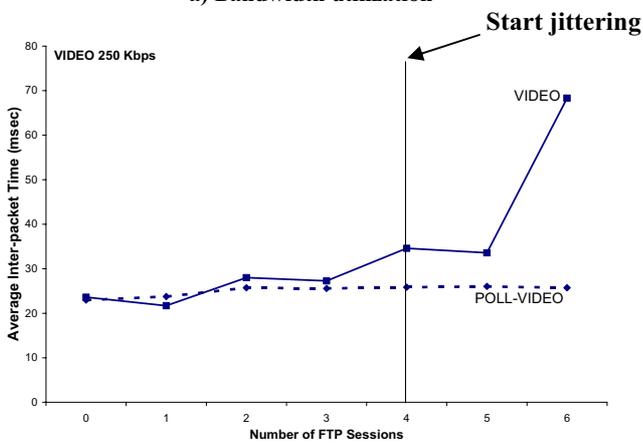

b) Average inter-packet time

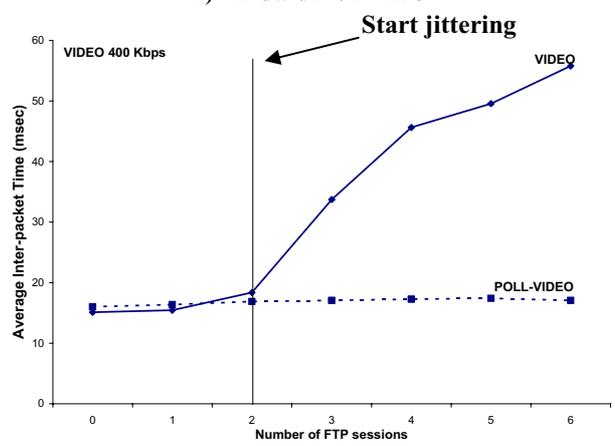

b) Average inter-packet time

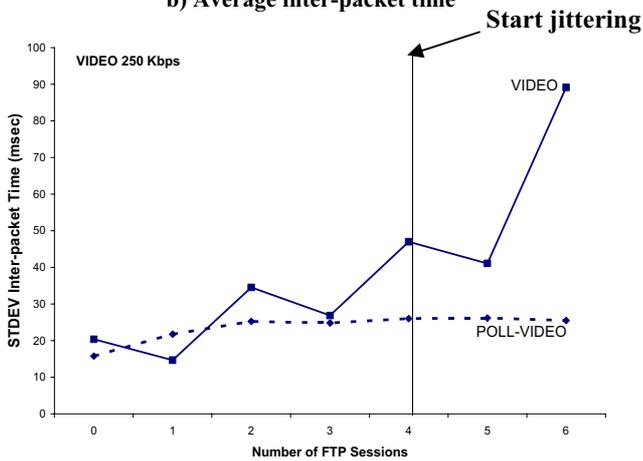

c) Standard deviation of the inter-packet time

Figure 5. Test results in case of 250 Kbps video stream.

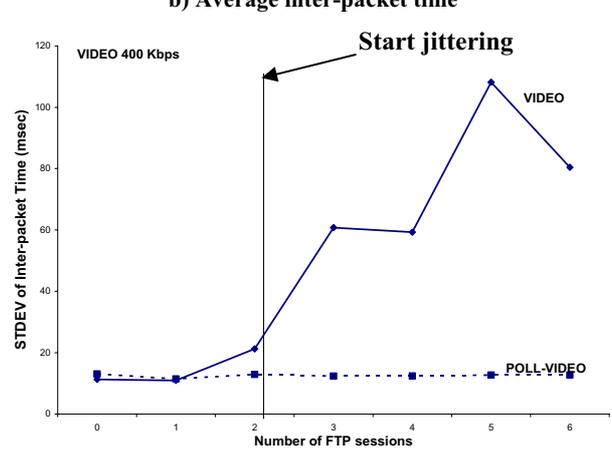

c) Standard deviation of the inter-packet time

Figure 6. Test results in case of 400 Kbps video stream.